\documentclass{appolb}
\usepackage{epsfig}
\usepackage{epsf,amsmath,amssymb,graphicx,scalefnt}
\usepackage{color}
\usepackage{rotating}

\newcommand{\abbrev}{\scalefont{.9}\rm}
\newcommand{\be}{\begin{equation}}
\newcommand{\ee}{\end{equation}}
\newcommand{\bea}{\begin{eqnarray}}
\newcommand{\eea}{\end{eqnarray}}
\newcommand{\lo}{{\abbrev LO}}
\newcommand{\nlo}{{\abbrev NLO}}
\newcommand{\nnlo}{{\abbrev NNLO}}
\newcommand{\mtop}{M_{t}}

\newcommand{\lhc}{{\abbrev LHC}}
\newcommand{\eqn}[1]{Eq.\,(\ref{#1})}

\newcommand{\order}[1]{{\cal O}(#1)}

\newcommand{\msbar}{\overline{\mbox{\abbrev MS}}}
\newcommand{\dd}{{\rm d}}

\begin{document}
\title{Top Mass Effects in Higgs Production at Hadron Colliders %
  \thanks{Presented by K.~J.~ Ozeren at the XXXIII International Conference of Theoretical Physics ``Matter to the Deepest'', Ustro\'n, Poland 11-16 September 2009.}%
}
\author{Kemal J. Ozeren
\address{Fachbereich C, Bergische Universit\"at Wuppertal\\42097 Wuppertal, Germany}
}
\maketitle
\begin{abstract}
We derive the first four terms of an expansion in $m_H^2/m_t^2$ of the total Higgs cross section through gluon fusion. At \nlo{} we demonstrate the excellent convergence of this series to the known result keeping the exact top mass dependence. At \nnlo{} there is no known exact result, and our work represents a thorough quantitative investigation of the effects of finite top mass at this order. We discuss the applicability of our approach, and present numerical results for the \lhc{} and Tevatron.
\end{abstract}
  
\section{Introduction}

\begin{figure}
\centering
\includegraphics[width=1.4in]{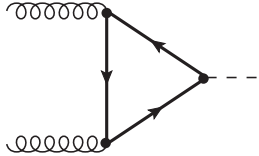}
\caption{Leading order diagram for the process $gg \to H$.}
\label{fig:lead}
\end{figure}

The forthcoming \lhc{} experiments are expected to elucidate the mechanism of electroweak symmetry breaking. The most popular theoretical models of this phenomenon invoke a new scalar field called the Higgs boson, and the \lhc{} has been designed with this firmly in mind. To discover the Higgs it is important to thoroughly understand its production and decay modes, as well as any relevant backgrounds. In this talk we describe recent work concerning the production of a Standard Model Higgs boson through gluon fusion. This process proceeds via a massive top quark, and as one can see from Fig.~\ref{fig:lead} the leading order diagrams are already one loop. At this order an exact analytical expression for the cross section is known, but at higher orders in $\alpha_s$ analytic formulae retaining the exact top mass dependence are only partially known \cite{Harlander:2005rq,Anastasiou:2006hc,Aglietti:2006tp}, although a numerical code known as {\abbrev HIGLU} \cite{Spira:1995mt} is available, which evaluates the \nlo{} cross section exactly. Fortunately, one finds that the cross section is very well reproduced  by working in an effective theory in which the top quark is integrated out, and then weighting with the leading order mass dependence. Comparisons with {\abbrev HIGLU} show that this procedure works extremely well up to \nlo{} \cite{Dawson:1990zj,Djouadi:1991tka,Spira:1995rr}. This is usually taken as sufficient justification for also using the effective theory approach at higher orders; the \nnlo{} contributions ~\cite{Harlander:2002wh,Anastasiou:2002yz,Ravindran:2003um} are known only in the effective theory.

To test the heavy-top limit approximation, we work in the full theory, including the top quark, and evaluate the cross section at \nnlo{} as an asymptotic series in $1/M_t$. The technology to perform this expansion is well known, and automatised in the {\tt q2e/exp} framework~\cite{Harlander:1997zb}. We assume $M_t$ heavier than all other scales in the problem, and express all Feynman integrals as convolutions of massive vacuum integrals with at most three loops, and massless 3/4/5-point functions through 2/1/0 loops, respectively. We note that the purely virtual contributions at \nnlo{} have recently been calculated by two separate groups~\cite{Harlander:2009bw,Pak:2009bx}.

We are able to directly evaluate the single real emission contribution in terms of hypergeometric functions, which we then expand in $\epsilon$ with the {\abbrev HypExp}~\cite{Huber:2007dx} package. For the double real emission part we perform an expansion in powers of $(1-x)$ and then integrate term by term. To demonstrate the cancellation of infrared poles we must then of course similarly expand the single real part. After renormalising the coupling $\alpha_s$ (for which we adopt the $\msbar{}$ scheme), top mass $M_t$ and gluon wave function (both in the on-shell scheme) we are left with only $1/\epsilon^2$ and $1/\epsilon$ infrared poles. These are absorbed in the {\abbrev PDF}s as per the usual mass factorisation procedure.

The description provided here is necessarily brief - we refer the reader to Ref.~\cite{Harlander:2009mq} for more details, full analytic results and comprehensive references.

\section{Small-x Behaviour}

The total cross section is a convolution of the partonic cross section with non-perturbative {\abbrev PDF}s. Equivalently, we can write it as an integral over the luminosity,
\begin{equation}
\sigma = \sum_{\alpha,\beta}\int_{m_H^2/S}^1 \dd x \,
    {\cal E}_{\alpha\beta}(x)\,\hat\sigma_{\alpha\beta}(x),
\label{eq::heavytop}
\end{equation}
where $x$ is related to the partonic centre of mass energy by $x = M_h^2/\hat{s}$. The small $x$ region of this integral therefore corresponds to high energy. For the \lhc{} the partonic cms energy can range up to 14~TeV, and so our assumption that the top mass is the heaviest scale in the process clearly breaks down here. Fortunately the behaviour of the cross section in the limit $x \to 0$ is known exactly, \ie keeping the full top mass dependence~\cite{Marzani:2008az}. We can therefore improve our results, which are valid for large $x$, by smoothly matching them onto the small-$x$ results of Ref.~\cite{Marzani:2008az}.

The small-$x$ results are,
\begin{equation}
\begin{split}
\hat\sigma^{(1)}_{gg}(x) &= 3\,\sigma_0\,{\cal C}^{(1)}+ \order{x}\,,\qquad
\hat\sigma^{(2)}_{gg}(x) = -9\,\sigma_0\,{\cal C}^{(2)}\ln x + c + \order{x}\,,
\label{eq::xto0}
\end{split}\end{equation}
for the \nlo{} and \nnlo{} cross sections respectively, where the coefficients ${\cal C}^{(1)}$ and ${\cal C}^{(2)}$ are given in Ref.~\cite{Marzani:2008az} in the form of a numerical table, out of which we construct simple interpolating functions. The constant $c$ was not determined, and we set it to zero. Our strategy for matching these expressions with our results is to construct functions which have the correct behaviour in each of the limits $x \to 0$ and  $x \to 1$, up to some order $N$ in an expansion in powers of $(1-x)$. We write
\begin{equation}
\begin{split}
\hat\sigma^{(1)}_{gg}(x) &= \hat\sigma^{(1),N}_{gg}(x) +
(1-x)^{N+1}\,\left[3\,\sigma_0 {\cal C}^{(1)} - \hat\sigma^{(1),N}_{gg}(0)
  \right]\,,\\ \hat\sigma^{(2)}_{gg}(x) &= \hat\sigma^{(2),N}_{gg}(x) -
9\,\sigma_0{\cal C}^{(2)}\left[ \ln x + \sum_{k=1}^N\frac{1}{k}(1-x)^k
  \right]\,,
\label{eq::match}
\end{split}
\end{equation}
where $\hat\sigma_{gg}^{(n),N}$ denotes the expansion of the partonic
cross section around $x=1$ through $\order{(1-x)^N}$.

\section{Numerical Results}

In order to strictly test the heavy-top limit, we apply a consistent $1/M_t$ expansion to the partonic cross section, without factoring the \lo{} mass dependence into the higher order terms. At \nlo{} we therefore define,
\be
\label{eq:NLO}
{\hat \sigma}^{\rm NLO}_{\alpha \beta}(M_t^n) = \sigma_ 0 \delta_{\alpha g} \delta_{\beta g}\delta(1-x) + \frac{\alpha_s}{\pi} {\hat \sigma}^{(1)}_{\alpha \beta}(M_t^n).
\ee

\begin{figure}
  \begin{center}
    \begin{tabular}{cc}
      \includegraphics[bb=110 265 465
        560,width=.41\textwidth]{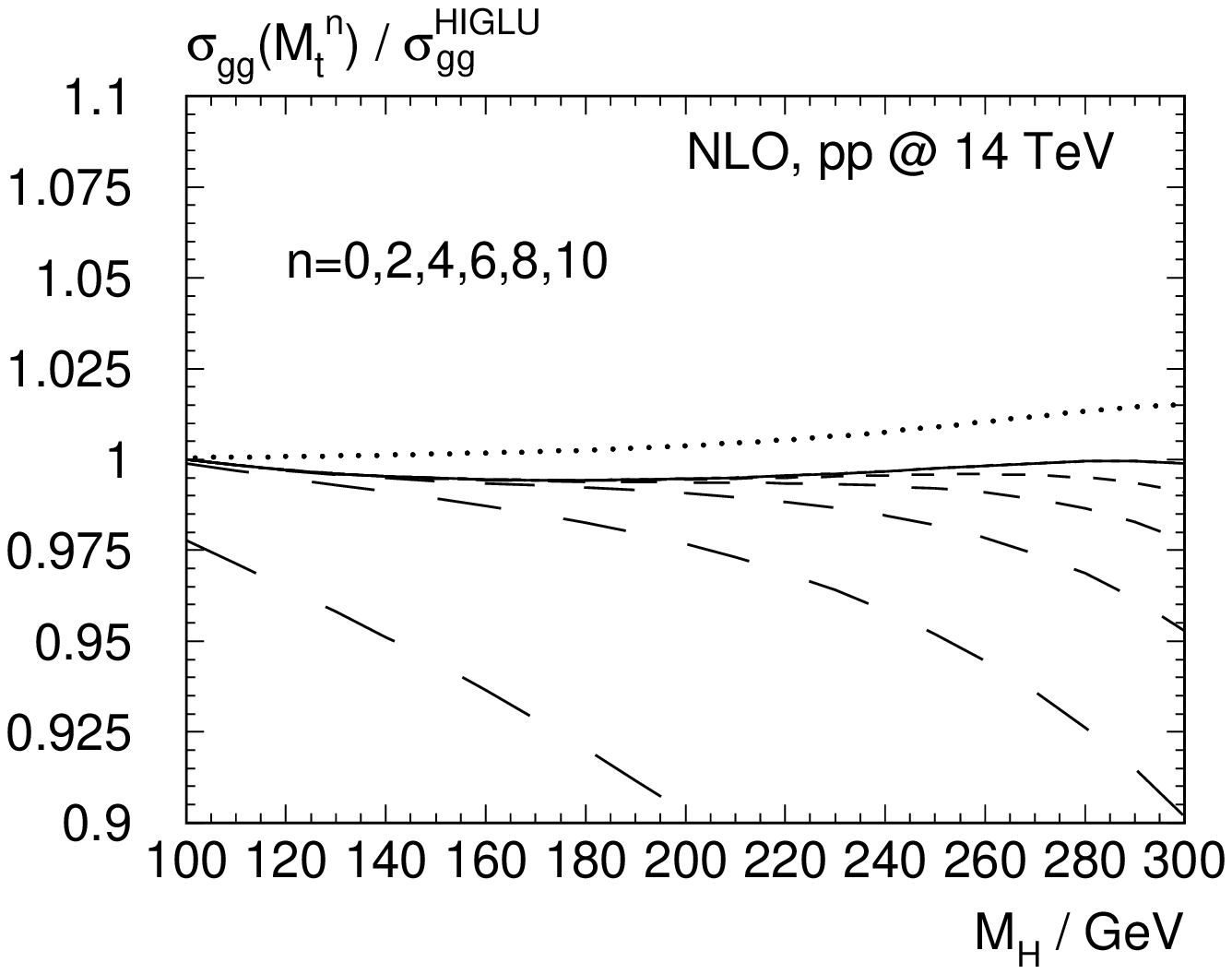} &
      \includegraphics[bb=110 265 465
        560,width=.41\textwidth]{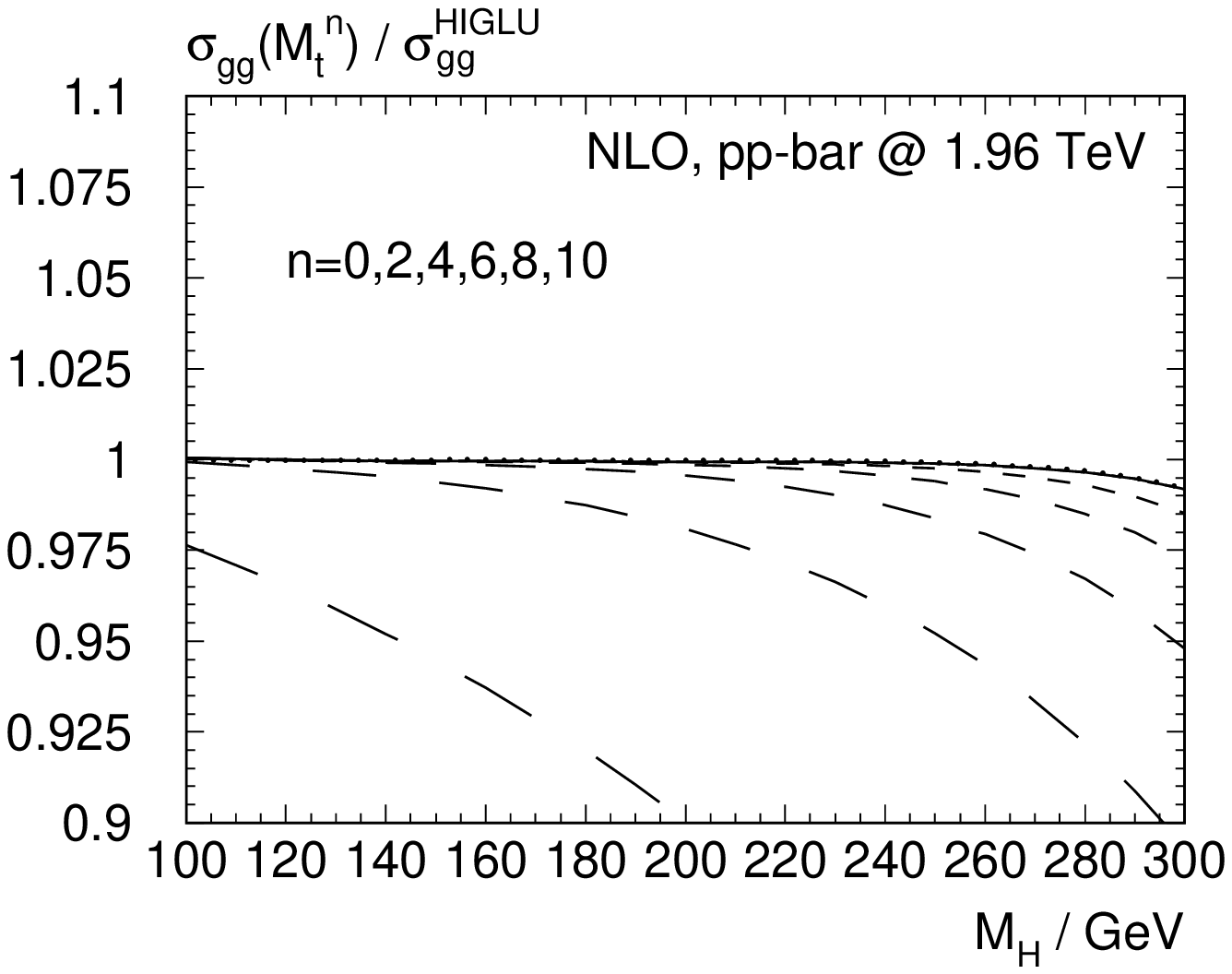}\\
      (a) & (b)
    \end{tabular}
    \parbox{.9\textwidth}{
      \caption[]{\label{fig:ratnloggmt}\sloppy Ratio of the $gg$
        induced component of the \nlo{} hadronic cross section as
        obtained from \eqn{eq::match} to the value obtained from {\tt
          HIGLU}~\cite{Spira:1995mt}, when keeping successively higher
        orders in $1/\mtop$ (decreasing dash-length corresponds to
        increasing order); the dotted line is the result obtained from
        the pure soft expansion $\hat\sigma_{gg}^{(1),N}$ through order
        $1/\mtop{}^{10}$ without the matching of \eqn{eq::match}.}}
  \end{center}
\end{figure}

In Fig.~\ref{fig:ratnloggmt} we compare the $gg$-channel \nlo{} cross section, evaluated according to \eqn{eq:NLO}, including successive terms in the $1/M_t$ expansion, with the exact result from the numerical code {\abbrev HIGLU}. We observe excellent convergence of the series at both the \lhc{} and Tevatron. Unfortunately the low-$x$ behaviour of the other subprocesses ($qg$, $q\bar q$ and also the \nnlo{} channels $qq$ and $qq'$) is not known. However, their numerical contribution at \nlo{} is small, at the level of a few percent in the case of $qg$ and at the permille level for $q\bar q$.

\begin{figure}
  \begin{center}
    \begin{tabular}{cc}
      \includegraphics[bb=110 265 465
        560,width=.41\textwidth]{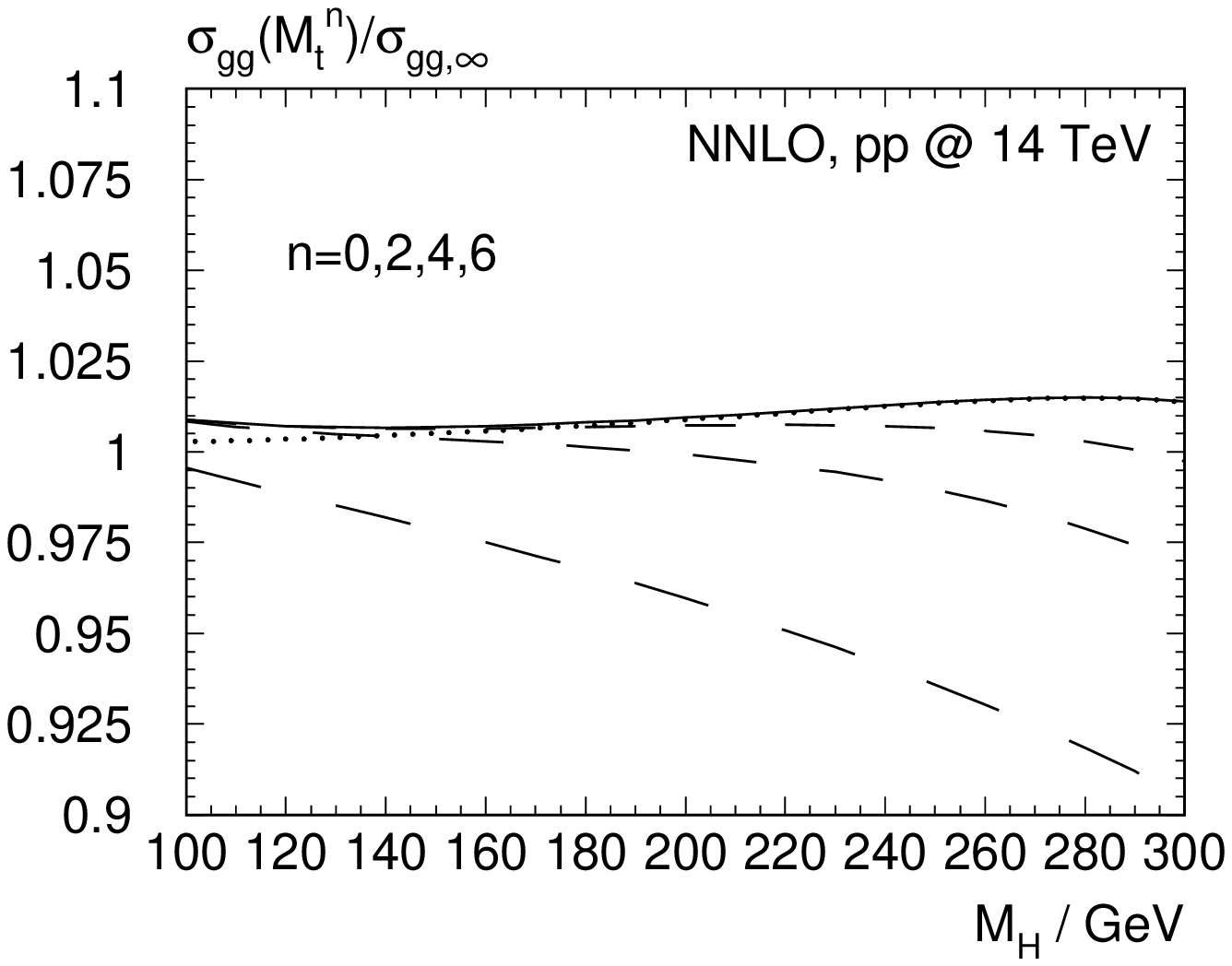} &
      \includegraphics[bb=110 265 465
        560,width=.41\textwidth]{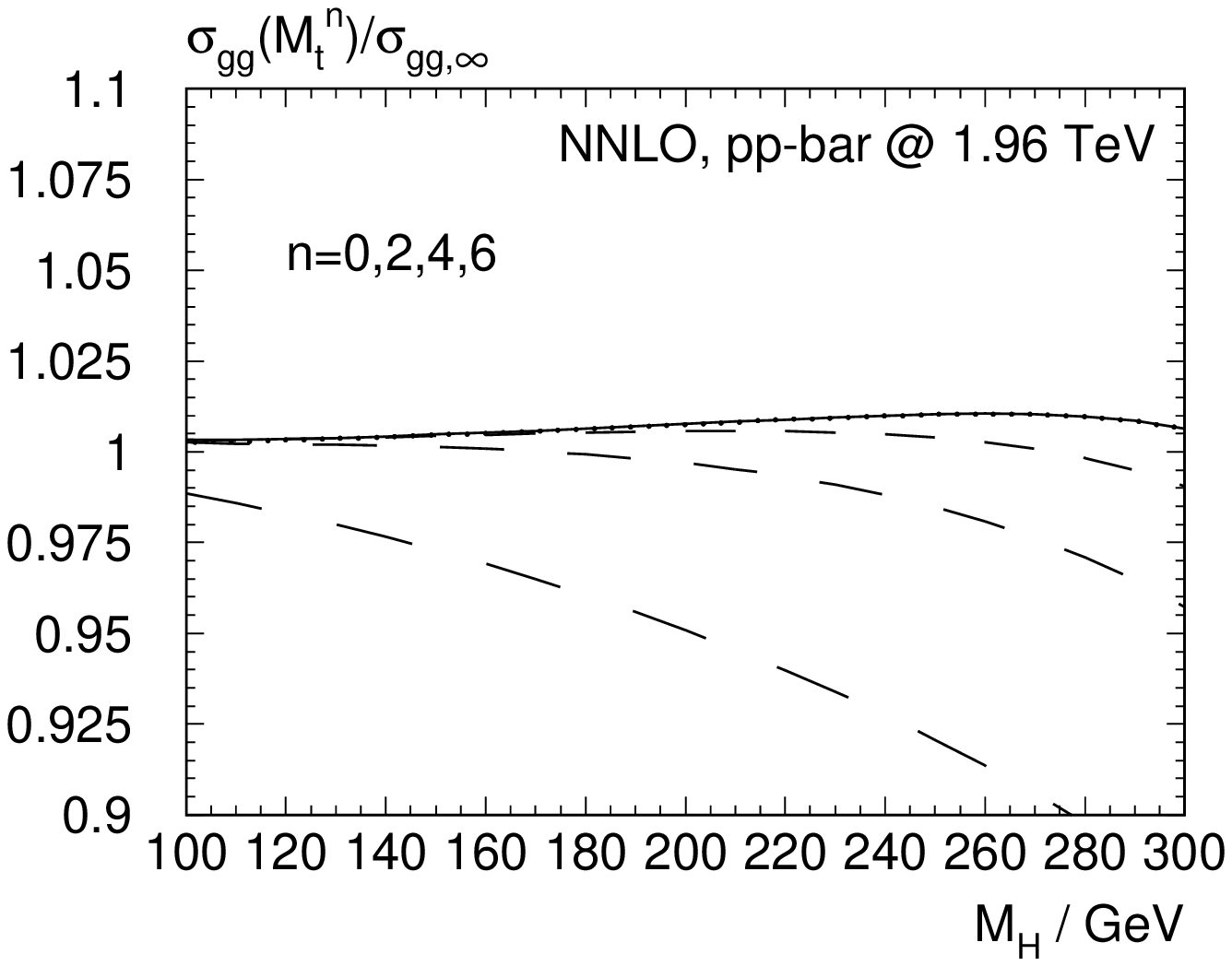}\\
      (a) & (b)
    \end{tabular}
    \parbox{.9\textwidth}{
      \caption[]{\label{fig:ratnnloggmt}\sloppy Ratio of the $gg$
        induced component of the \nnlo{} hadronic cross section as
        obtained from \eqn{eq::match} to the heavy-top result of
        \eqn{eq::heavytop} (decreasing dash-length corresponds to
        increasing order in $1/\mtop{}$); the dotted line is the result
        obtained from the pure soft expansion $\hat\sigma_{gg}^{(2),N}$
        through order $1/\mtop{}^6$ without the matching of
        \eqn{eq::match}.}}
  \end{center}
\end{figure}

The success of our approach at \nlo{} means we can confidently apply it at higher orders. In Fig.~\ref{fig:ratnnloggmt} we compare the \nnlo{} total cross section in our $1/M_t$ expansion approach with that obtained in the heavy top effective theory. The \lo{} mass dependence is factored in only up to \nlo{}, so that we can examine the truly \nnlo{} mass effects. We observe very good convergence towards the heavy-top results, which assures us of the high quality of the latter. The two results differ by less that $0.5\%$, which clearly justifies the use of the heavy-top effective theory so far in the literature, and also in future studies.

\def\app#1#2#3{{\it Act.~Phys.~Pol.~}\jref{\bf B #1}{#2}{#3}}
\def\apa#1#2#3{{\it Act.~Phys.~Austr.~}\jref{\bf#1}{#2}{#3}}
\def\annphys#1#2#3{{\it Ann.~Phys.~}\jref{\bf #1}{#2}{#3}}
\def\cmp#1#2#3{{\it Comm.~Math.~Phys.~}\jref{\bf #1}{#2}{#3}}
\def\cpc#1#2#3{{\it Comp.~Phys.~Commun.~}\jref{\bf #1}{#2}{#3}}
\def\epjc#1#2#3{{\it Eur.\ Phys.\ J.\ }\jref{\bf C #1}{#2}{#3}}
\def\fortp#1#2#3{{\it Fortschr.~Phys.~}\jref{\bf#1}{#2}{#3}}
\def\ijmpc#1#2#3{{\it Int.~J.~Mod.~Phys.~}\jref{\bf C #1}{#2}{#3}}
\def\ijmpa#1#2#3{{\it Int.~J.~Mod.~Phys.~}\jref{\bf A #1}{#2}{#3}}
\def\jcp#1#2#3{{\it J.~Comp.~Phys.~}\jref{\bf #1}{#2}{#3}}
\def\jetp#1#2#3{{\it JETP~Lett.~}\jref{\bf #1}{#2}{#3}}
\def\jphysg#1#2#3{{\small\it J.~Phys.~G~}\jref{\bf #1}{#2}{#3}}
\def\jhep#1#2#3{{\small\it JHEP~}\jref{\bf #1}{#2}{#3}}
\def\mpl#1#2#3{{\it Mod.~Phys.~Lett.~}\jref{\bf A #1}{#2}{#3}}
\def\nima#1#2#3{{\it Nucl.~Inst.~Meth.~}\jref{\bf A #1}{#2}{#3}}
\def\npb#1#2#3{{\it Nucl.~Phys.~}\jref{\bf B #1}{#2}{#3}}
\def\nca#1#2#3{{\it Nuovo~Cim.~}\jref{\bf #1A}{#2}{#3}}
\def\plb#1#2#3{{\it Phys.~Lett.~}\jref{\bf B #1}{#2}{#3}}
\def\prc#1#2#3{{\it Phys.~Reports }\jref{\bf #1}{#2}{#3}}
\def\prd#1#2#3{{\it Phys.~Rev.~}\jref{\bf D #1}{#2}{#3}}
\def\pR#1#2#3{{\it Phys.~Rev.~}\jref{\bf #1}{#2}{#3}}
\def\prl#1#2#3{{\it Phys.~Rev.~Lett.~}\jref{\bf #1}{#2}{#3}}
\def\pr#1#2#3{{\it Phys.~Reports }\jref{\bf #1}{#2}{#3}}
\def\ptp#1#2#3{{\it Prog.~Theor.~Phys.~}\jref{\bf #1}{#2}{#3}}
\def\ppnp#1#2#3{{\it Prog.~Part.~Nucl.~Phys.~}\jref{\bf #1}{#2}{#3}}
\def\rmp#1#2#3{{\it Rev.~Mod.~Phys.~}\jref{\bf #1}{#2}{#3}}
\def\sovnp#1#2#3{{\it Sov.~J.~Nucl.~Phys.~}\jref{\bf #1}{#2}{#3}}
\def\sovus#1#2#3{{\it Sov.~Phys.~Usp.~}\jref{\bf #1}{#2}{#3}}
\def\tmf#1#2#3{{\it Teor.~Mat.~Fiz.~}\jref{\bf #1}{#2}{#3}}
\def\tmp#1#2#3{{\it Theor.~Math.~Phys.~}\jref{\bf #1}{#2}{#3}}
\def\yadfiz#1#2#3{{\it Yad.~Fiz.~}\jref{\bf #1}{#2}{#3}}
\def\zpc#1#2#3{{\it Z.~Phys.~}\jref{\bf C #1}{#2}{#3}}
\def\ibid#1#2#3{{ibid.~}\jref{\bf #1}{#2}{#3}}
\def\otherjournal#1#2#3#4{{\it #1}\jref{\bf #2}{#3}{#4}}
\newcommand{\jref}[3]{{\bf #1} (#2) #3}
\newcommand{\bibentry}[4]{#1, {\it #2}, #3\ifthenelse{\equal{#4}{}}{}{, }#4.}
\newcommand{\hepph}[1]{{\tt hep-ph/#1}}
\newcommand{\mathph}[1]{[math-ph/#1]}
\newcommand{\arxiv}[2]{{\tt arXiv:#1}}

\end{document}